\documentclass{article}
\usepackage{spconf,amsmath,graphicx,subfigure,enumitem}
\usepackage{nicefrac}
\usepackage[table]{xcolor}
\usepackage{hyperref}
\title{Audio-based identification of beehive states} 

\name{\begin{tabular}{c} In\^{e}s Nolasco{$^2$}, Alessandro Terenzi{$^1$}, Stefania Cecchi{$^1$}, \\ Simone Orcioni{$^1$}, Helen L. Bear{$^2$}, and  Emmanouil Benetos{$^{2,3}$} \thanks{This work was supported by a Universit\'a Politecnica delle Marche Research Grant, EPSRC Grant EP/R01891X/1, RAEng Research Fellowship RF/128, and by The Alan Turing Institute under EPSRC grant EP/N510129/1.}\end{tabular}}
\address{$^1$Department of Information Engineering, Universit\'a Politecnica delle Marche, Italy \\
$^2$ School of EECS, Queen Mary University of London, UK\ \ \ \ $^3$ The Alan Turing Institute, UK}
\begin{document}
\ninept
\maketitle
\begin{abstract}
The absence of the queen in a beehive is a very strong indicator of the need for beekeeper intervention. Manually searching for the queen is an arduous recurrent task for beekeepers that disrupts the normal life cycle of the beehive and can be a source of stress for bees. Sound is an indicator for signalling different states of the beehive, including the absence of the queen bee. In this work, we apply machine learning methods to automatically recognise different states in a beehive using audio as input. % The system is built on top of a method for beehive sound recognition in order to detect bee sounds from other external sounds.
We investigate both support vector machines and convolutional neural networks for beehive state recognition, using audio data of beehives collected from the NU-Hive project. Results indicate the potential of machine learning methods as well as the challenges of generalizing the system to new hives.
\end{abstract}
\section{Introduction}
\label{sec:intro}

Among insects, honey bees (\textit{Apis mellifera L.}) are well known for their positive effects. Their importance is not limited to the production of honey, beeswax, royal jelly, and propolis but they are also at the basis of plant pollination, playing a key role in the proliferation of both spontaneous and cultivated flora. In recent years multiple stress factors have led to a decline of honey bee colonies \cite{Cane2007} and this event has emphasized the significance of a continuous and extensive monitoring to investigate factors that may negatively affect the life cycle of bees. 

In this context, the analysis of sound generated within the bee hives is an important approach for non-invasive monitoring \cite{ntalampiras2012acoustic}. Vibration and sound signals are used by honey bees to communicate within the colony \cite{Frings:1957,Michelsen:1986}. Honey bees produce their sounds by means of gross body movements, wing movements, high-frequency muscle contractions without wing movements, and pressing the thorax against the substrates or another bee \cite{Kirchner1993,Hrncir2006,Hunt2013}. 

%One of the first attempts of exploiting the beehive sound was done with the Apidictor \cite{woods1957}, an electronic system aimed to detect swarming, making use of analogue low pass filters.  COMMENTED OUT not directly relevant , traded for space. 
In recent years, several studies have underlined that some behaviors of the honey bees are strictly related to variation in produced sound \cite{Kirchner1993,Hunt2013,Cejrowski2018,robles2018}. In particular, these works have proved that there is a strict correlation between the amplitudes and frequencies of the bee hive sounds and some events like swarming \cite{Dietlein:1985,Ferrari:2006,Ferrari:2008,quandour2014} and queen presence \cite{Cejrowski2018,Hunt2013,robles2018}. In \cite{Bromenshenk:2007}, a relation between sound and changes in environmental conditions have been reported. Furthermore, it has been demonstrated that sound analysis could be a powerful instrument in pest monitoring e.g., in \cite{quandour2014} it has been used for varroa-mite detection.

Recent works in beehive sound analysis are carried out through a computational bioacoustic scene analysis perspective \cite{chapter11_book}.
In this context, relevant representations for these audio signals are combined with machine learning methods in order to develop systems that can automatically distinguish between different states of a hive. In \cite{Amlathe2018} and \cite{Inolasco_dissertation}, the authors explore the use of Mel spectra and Mel-frequency cepstral coefficients (MFCCs) together with different machine leaning methods to detect hives with and without the queen bee.
In the context of computational sound scene analysis research, state-of-the-art methods for sound scene recognition as \cite{dcase2017web_acousticScene_results} show, are mainly based in Convolutional Neural Networks (CNNs). In \cite{inolasco_DCASE2018}, the authors explore the use of CNNs to the problem of beehive sound identification and highlight the long-term aspects of such sounds. They stress the need for long-term contextual representations for modeling such data. Also, in \cite {Bisot2017}, the authors present a method to extract long-term features from spectrograms for the task of sound scene classification.

This work expands the preliminary work of \cite{Inolasco_dissertation} to investigate the potential of traditional and neural network-based machine learning methods exploiting MFCCs, Mel spectrograms, and the Hilbert Huang Transform (HHT) \cite{huang2005} as features to determine the presence of the queen bee in a hive. 

% ines- changed order of novelty so more enphasis can be give to the application of deep learning methods
The novelty of the proposed approach is related not only to the application of deep learning methods to this problem, but also to the use of HHT as spectral representations and the design of representations suitable for modelling long-term temporal context.
%to the use of HHT as spectral representations and the design of representations suitable for modelling long-term temporal context, but also by exploring %the use in combination of the HHT with MFCCs and Mel spectra, and we app
%the application of deep learning methods to this specific problem. 
The novelty is underlined in the experimental results where the proposed system has been tested on real audio data collected from the NU-Hive \cite{cecchi2018a} project. By designing a \emph{hive-independent} evaluation setup, inspired by speaker-independent evaluations in speech processing \cite{burton2018}, we demonstrate the validity and potential of the developed system in a real-world scenario.

%- 4th paragraph: "The novelty of the proposed work is related to the use of HHT for spectral feature extraction in combination with the MFCC." -> only? I think that there are several more novel aspects, including the use of deep learning methodologies for this task for the first time. Also this work is using mel spectra, not just MFCCs. Note that there is also novelty in creating a CNN architecture that combines Mel spectrograms and the HHT feature. There is also novelty in designing representations suitable for modelling long-term temporal context. Note that you are not competing with respect to novelty with your MSc thesis,which can be positioned as a pilot study that led to this paper.- 4th paragraph: you can also discuss the aspect of creating a system that is able to generalize to unseen hives (you can also mention the "hive-independent" term, mirroring "speaker-independent" evaluations done in speech processing).- 4th paragraph: this part needs to be more explicit on the new aspects of this paper compared to previous work. As before, there is no competition with your MSc thesis, which can be positioned as related work. You can also compare this paper with the MSc thesis of Amlathe - what is new in this paper?

The paper is organized as follows. Section~\ref{sec:prop} describes the proposed approach, including the feature extraction and classification methodologies used. Section~\ref{sec:results} presents the data acquired in honey bee hives, the experimental setup, and the obtained results. Finally, conclusions and future directions are reported in Section~\ref{sec:conc}.

\section{Proposed approach}
\label{sec:prop}

%overall description of the method starting from a general scheme
The proposed approach is based on two steps. Firstly, feature extraction is carried out using MFCCs, Mel spectrograms, and the HHT algorithm, with the aim of determining the frequency behaviour of the beehive when the queen is present or not. Secondly, classification of beehive states is achieved using both support vector machines (SVMs) and convolutional neural networks (CNNs), with different combinations of features and parameters as appropriate. 

\begin{figure}[t]
  \centering
  %\includegraphics[scale=0.41]{./img/mfcc_schema.eps}\\
  %\vspace{0.3cm}
  \includegraphics[scale=0.41]{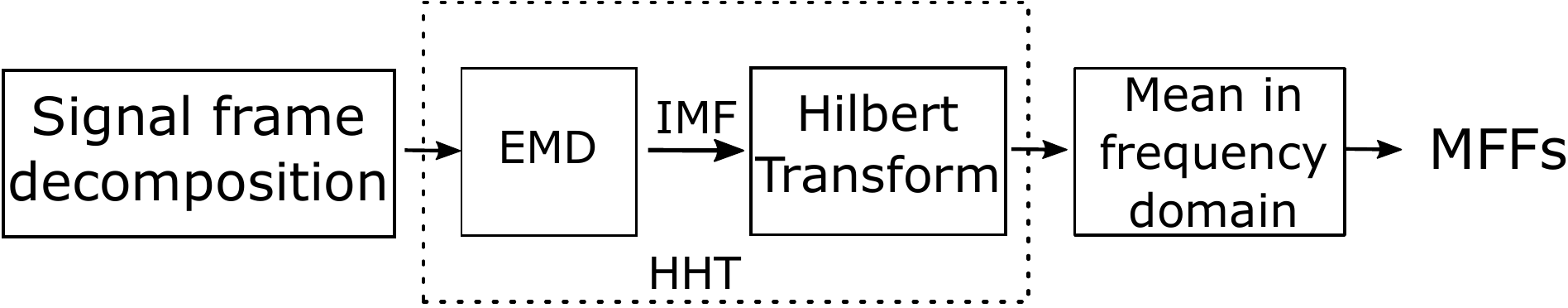}
 \caption{Feature extraction procedure based on HHT.}\label{fig:EMD_schema}
 \vspace{-3mm}
\end{figure}

\begin{figure*}
    \centering
    \subfigure[\label{subfig:mel}]{\includegraphics[width=5.6cm]{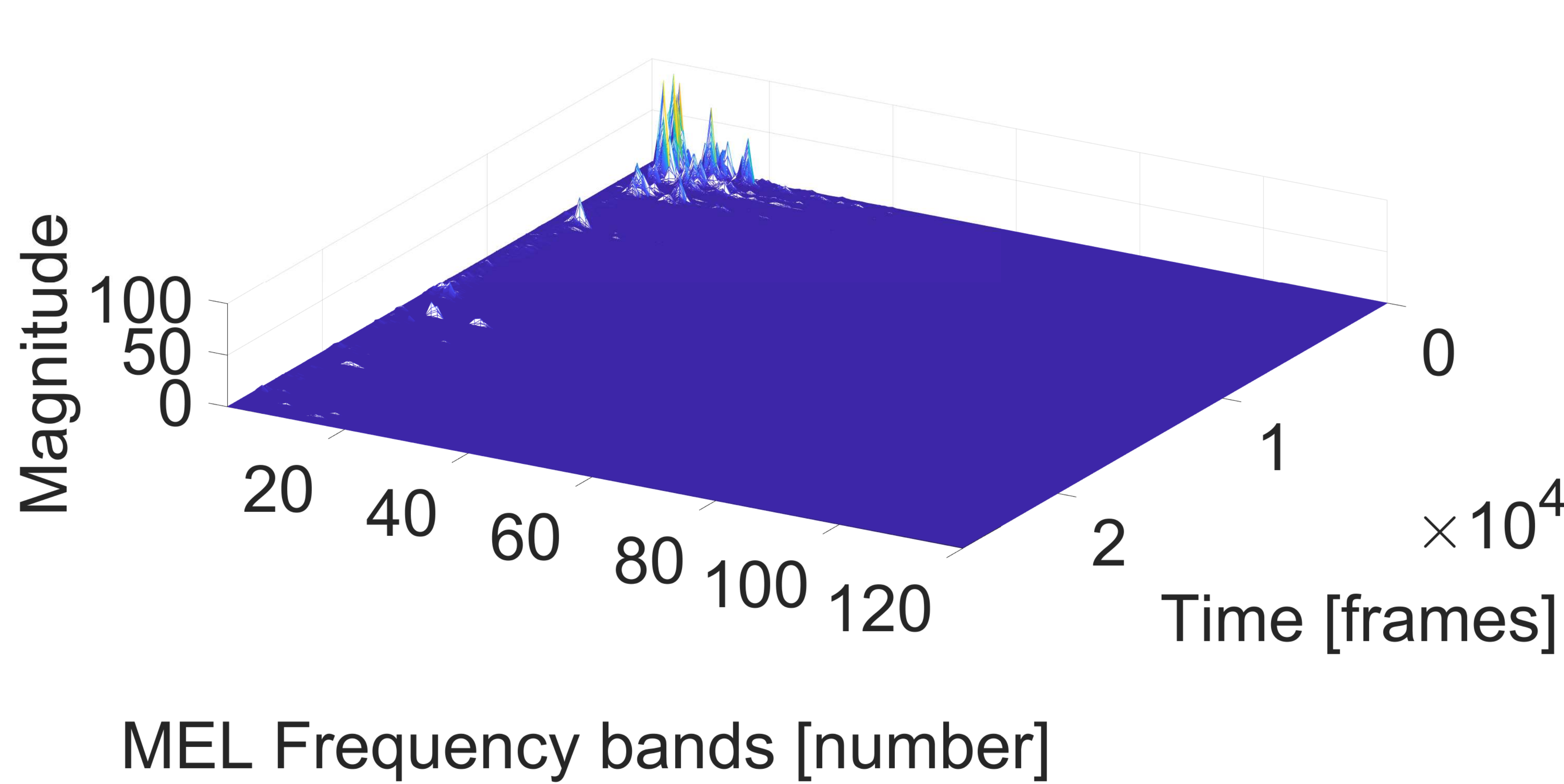}}
    \subfigure[\label{subfig:mfcc}]{\includegraphics[width=5.6cm]{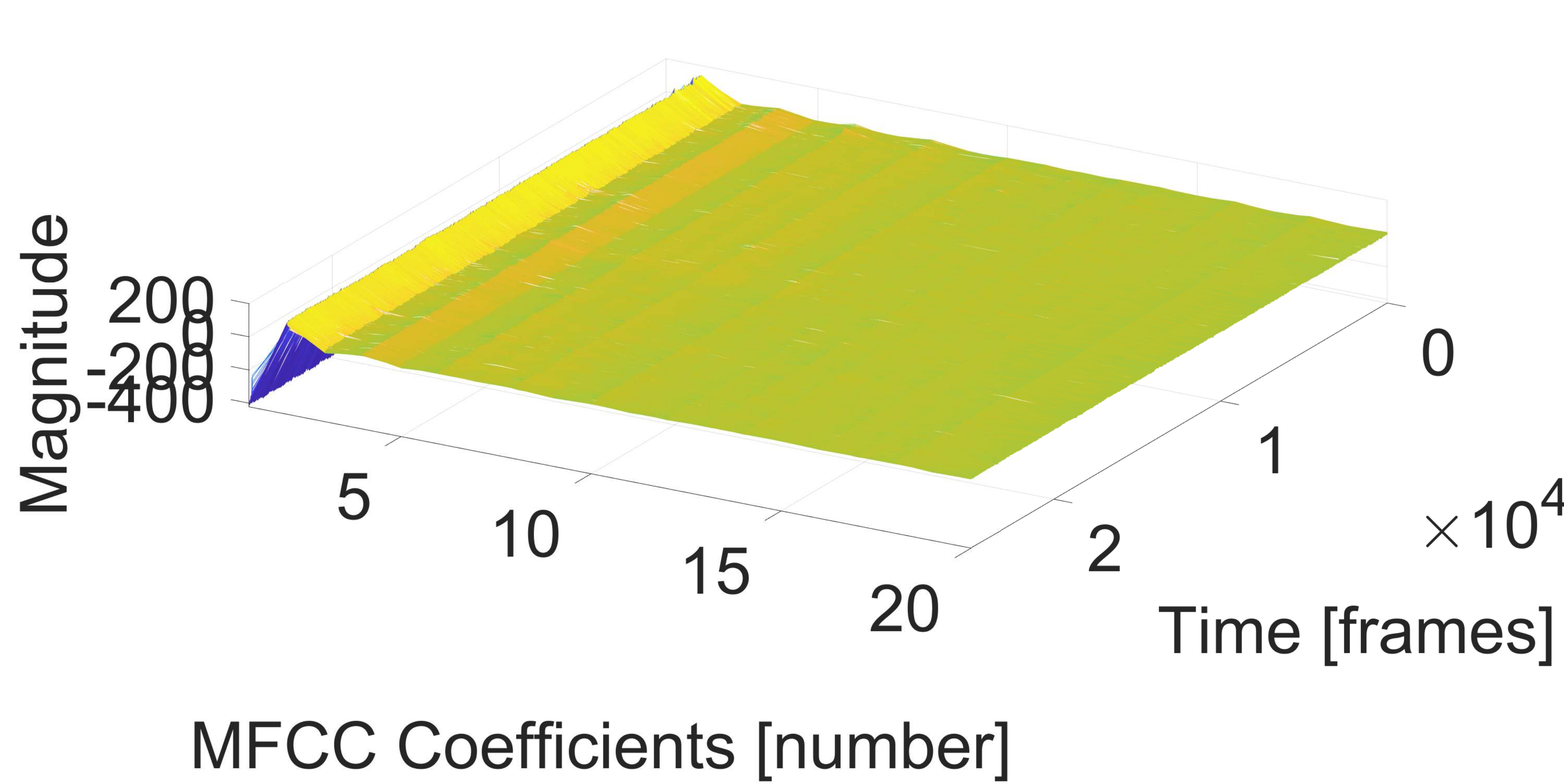}}
    \subfigure[\label{subfig:hht}]{\includegraphics[width=5.6cm]{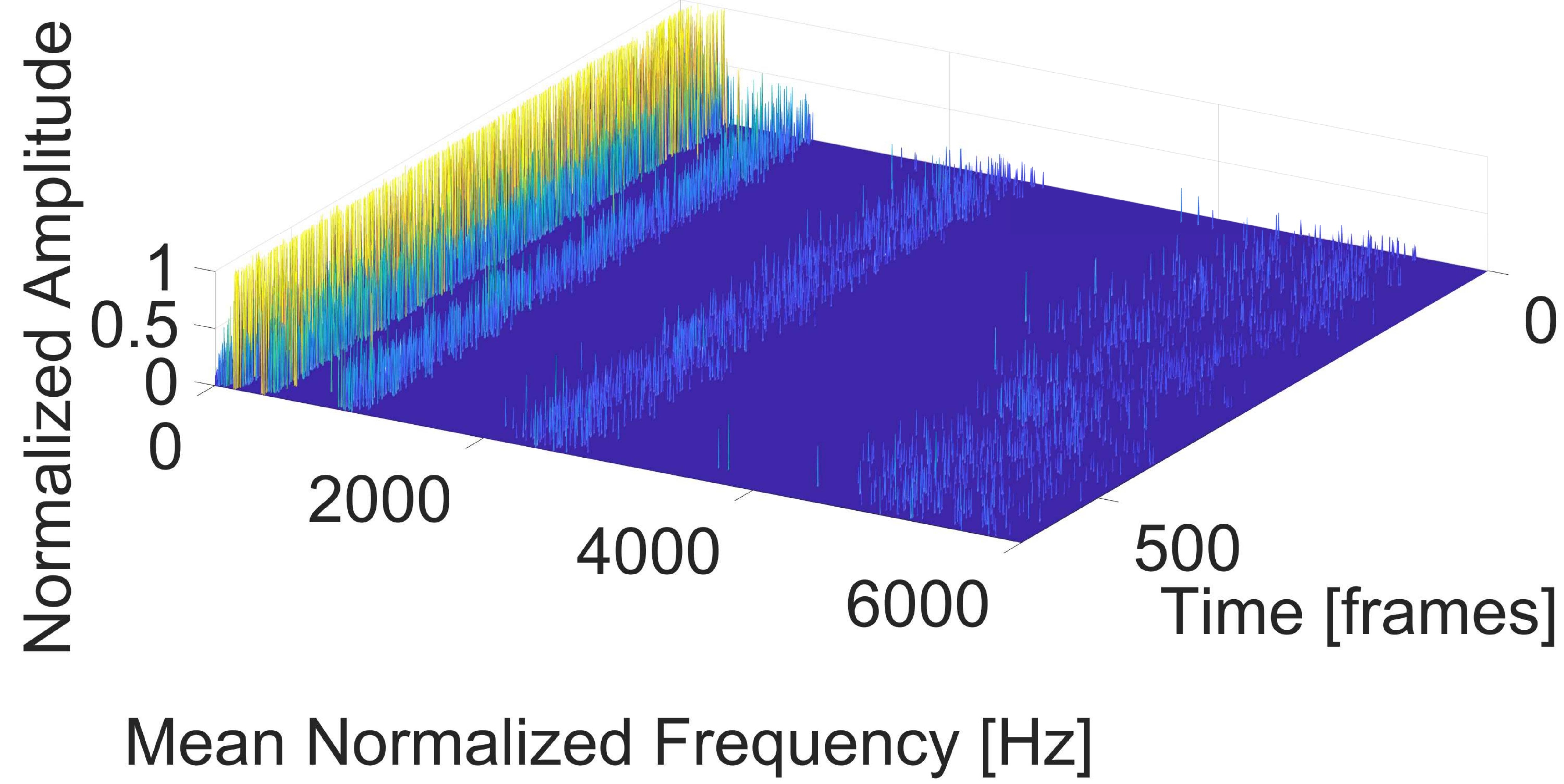}}
   \caption{Comparison of feature extraction: \subref{subfig:mel} Mel spectra considering the Mel frequency bands, \subref{subfig:mfcc} MFCCs considering the obtained coefficients, and \subref{subfig:hht} HHT-based features considering the extracted mean normalized frequency as function of magnitude/amplitude and time.}
    \label{fig:feat}
    \vspace{-3mm}
\end{figure*}

\subsection{Feature extraction}\label{sec:emd}
Three types of features are extracted to use within the classifiers. 
First, both Mel spectra and MFCCs, which are commonly used representations in the context of computational sound scene analysis \cite{chapter4_book}, are used in this work. We compute MFCCs with 20 coefficients and Mel spectra with 120 frequency bands.

%MFCCs \cite{serizel2018acoustic} are extracted from the audio recording using 20 coefficients
%Also Mel spectra \cite{} with 120 band is obtained 
%MFCC description
% they are calculated with the following steps. The Fourier transform of the windowed sound signal is firstly computed to derive the magnitude spectrum, whereupon the result is mapped onto the Mel scale using a filterbank composed of triangularly shaped filters; then, MFCC coefficients are calculated as the discrete cosine transform of the logarithm of the spectrum at each of the Mel frequencies.

%- Fig. 1: as discussed, this figure can lose the MFCC step (as it's common knowledge in the ICASSP community), and can focus only on the HHT feature extraction.- 2nd paragraph: the MFCC description can be shortened a lot. You also need to include a brief description of the mel spectrograms used for the CNNs (make sure to mention the number of MFCC coefficients and the number of Mel bands used in the two respective features).

%HHT description
The extraction of a new feature based on the HHT \cite{huang2005} is considered in the proposed procedure. This choice derives from the fact that it is known from the literature that honey bee sounds are non-stationary signals composed by a superimposition of tones at various frequencies \cite{quandour2014}.
Figure \ref{fig:EMD_schema} shows each step of the procedure. The signal is divided in frames of length $N=32000$ samples using a frequency sampling $f_s=32$~kHz, with analysis performed every 1 second.
% %%S: The signal is divided in frames of one second length that, using a frequency sampling $f_s=32$~kHz, result in $N=32000$ samples.
Each signal frame is decomposed with Empirical Mode Decomposition (EMD) \cite{Huang903} to obtain a set of basis functions which then are analyzed with the Hilbert transform. EMD uses adaptive basis functions for signal decomposition and this is useful when measured data are non-stationary and non-linear. The decomposition is based on the idea that the signals are composed of simple intrinsic modes of oscillations. Each intrinsic mode represents a simple oscillation with the same number of extrema and zero crossings. Each oscillation is represented by an Intrinsic Mode Function (IMF), which are the EMD basis functions and are defined by the following properties:
(A) The number of extrema and the number of zero-crossings must be either equal or differ at most by one. (B) The mean value of the envelope defined by the local maxima and the local minima is zero.

%\begin{itemize}[noitemsep]
%    \item The number of extrema and the number of zero-crossings must be either equal or differ at most by one.
 %   \item The mean value of the envelope defined by the local maxima and the local minima is zero. 
%\end{itemize}

Starting from this definition, any signal frame $x(n)$ with $n=1,\ldots,N$, can be decomposed following these steps:
\begin{enumerate}[noitemsep]
    \item Identify all local extrema.
    \item Connect all the local maxima by a cubic spline.
    \item Repeat the procedure to produce the lower envelope.
    \item Estimate their mean $m_1(n)$.
    \item The first estimation of the IMF can now be written as \\ $h_1(n)=x(n)-m_1(n)$.
    \item Repeat the procedure up to {$k$} times until the function \\
    $h_{1k}(n)=h_{1(k-1)}(n)-m_{1k}(n)$  \\
    does not satisfy the IMF properties.
    \item Now the first IMF component is equal to $c_1(n)=h_{1k}(n)$.
    \item Remove from the original signal the component $c_1(n)$ obtaining the first residue $r_1(n)=x(n)-c_1(n)$.
    \item Treat $r_1(n)$ as the new signal for the decomposition procedure.
\end{enumerate}
The process is stopped when the residue $r_n(n)$ becomes a monotonic function, or when the amplitude is less then a predetermined value. The number of components $c_j(n)$ is $M$ and it is not a predetermined value since it depends on the complexity of the signal.
Therefore, the original signal $x(n)$ can be reconstructed as a superimposition of estimated IMF plus the residue, i.e., 
\begin{equation}
\label{formula:emd1}
    x(n)=\sum\limits_{j=1}^M c_j(n)+r_n(n).
    \vspace{-4mm}
\end{equation}
In the specific case of honey bee sounds, we have empirically found that it is possible to obtain the original signal setting $M=10$.

After the EMD decomposition, the Hilbert Transform \cite{marple1999} is applied to each IMF and used for the estimation of the analytic signal $a_j(n)$ as follows:
\begin{equation}\label{formula:hilbert1}
    a_j(n)=c_j(n)+j\mathcal{H}\{c_j(n)\}
\end{equation}
where $\mathcal{H}$ indicates the Hilbert transform and $j=1,\ldots,M$. Then, equation (\ref{formula:hilbert1}) can be expressed in polar coordinates, i.e.,  $a_j(n)=A_j(n)e^{i\phi_j(n)}$
%\begin{equation}\label{formula:hilbert2}
%    a_j(n)=A_j(n)e^{i\phi_j(n)}
%\end{equation}
where $A_j(n)$ is the instantaneous amplitude of the signal, and $\phi_j(n)$ is the phase from which can be derived the instantaneous frequency $f_j(n)= \frac{f_s}{2\pi}\big[ \phi_j(n+1) - \phi_j(n) \big]$.
%\begin{equation}
%f_j(n)=\sum\limits_{n=1}^{N-1} \frac{1}{2\pi}\big[ \phi_j(n+1) - \phi_j(n) \big]
%\end{equation}
%Finally, the spectral feature is derived as a combination of the values obtained in Eq.(\ref{formula:hilbert1}) and (\ref{formula:hilbert2}) over the frame length $N$, i.e.,
%
Finally, the spectral features are derived considering the mean normalized frequencies (MNF) calculated as in \cite{Abdelouahad20018} and the amplitude calculated as a mean of all instantaneous amplitude, obtaining a spectrogram spanning over 10 min.
%the sum of the amplitudes corresponding to each frequency.
%Then, the mean frequency feature (MFF) is derived as follows:
%\begin{equation}\label{formula:mean}
%    \text{MFF}_j=\sum\limits_{n=1}^N A_j^2(n) f_j(n) / \sum\limits_{n=1}^N A_j^2(n).
%\end{equation}
%where $j=1,\ldots,M$ and $M$ is the number of components in the analyzed frame. This type of evaluation was used before in biomedical signal  analysis \cite{Abdelouahad20018}.

Fig.~\ref{fig:feat} compares the three extracted features in relation to a queenless hive state for a time interval of 10 minutes. It is evident how the HHT-based method is capable of expressing the frequency behaviour of the analyzed bee hives. 

%\begin{figure}[!t]
%    \centering
%    \subfigure[]{\includegraphics[width=8cm]{./img/mel.eps}}
%     \subfigure[]{\includegraphics[width=8cm]{./img/mfcc.eps}}
%      \subfigure[]{\includegraphics[width=8cm]{./img/hht.eps}}
%   \caption{Caption}
%    \label{fig:my_label}
%\end{figure}

%- Section 2.1: At this point it would be useful to display the HHT feature for an audio example, possibly along with a Mel spectrogram for comparison.- For the HHT feature extraction process, were specific parameter values used (e.g. number of basis functions? Hilbert transform settings?) which are useful for reproducing the exact features used in this paper?

\subsection{Classification} \label{sec:ml}

%%SVM 
For classification, a first approach employing SVMs and various feature combinations is carried out (Section~\ref{sec:SVMexperiments}).  %An SVM is a rather simple but powerful method that discriminates data in its classes based by finding a boundary that maximizes the distance between samples from the different classes. An important aspect of SVMs is the possibility of performing this procedure for data that is not linearly separable, through means of a kernel. \cite{} ... \textcolor{red}{REVIEW}
As indicated in \cite{Inolasco_dissertation}, SVMs are good classifiers for this problem when used with a radial basis function (RBF) kernel. 
Here, all SVMs are computed with the RBF kernel, with penalty parameter (C) of 1 and the gamma parameter of 1/(number of features). The input data consists of various combinations of features which are extracted from 10min audio recordings. The samples are normalized using z-score normalization across each training and test sets as described in Section~\ref{sec:expSetup}. %http://scikit-learn.org/stable/auto_examples/svm/plot_rbf_parameters.html

%CNN
Recent research in the field of computational sound scene analysis shows a clear dominance of data-driven deep learning methods such as CNNs over other traditional machine learning methods \cite{dcase2017web_acousticScene_results}. %  CNNs are an important and very powerful methods to extract relevant patterns from a rather raw image-based audio representation such as spectrograms. %By applying several convolution steps to the input image, they are able to learn the most discriminative aspects of the data in hierarchical levels of abstraction and relevance. 
Therefore, we adopt a CNN classifier designed to further explore how applicable these models are for beehive state recognition.  

When using CNNs it is important to consider the amount of data needed to train such large networks. To meet these constraints, we segment the original 10min audio samples into 1 min segments. Feature extraction is performed on this new set of shorter samples. To mitigate the loss of temporal context that the shorter segments bring and which was deemed important in \cite{inolasco_DCASE2018} for representing beehive sounds, we adapt a procedure to obtain long-term features introduced in \cite{Bisot2017}, where each spectrogram is transformed in a stack of averaged slices over time.

%data augmentation
To increase the generalization of the models, we carry out data augmentation on the dataset by creating 3 versions of each sample with a random pitch shift between -1 and 1 semitone.
%Normalization
The normalization of the data is performed frequency-wise with z-score normalization along the training set samples. The normalization parameters computed for the training set are then used to apply the same transformation to both validation and test sets.

The general network architecture, presented in Table~\ref{table:CNN_spectra_architecture}, consists of four convolutional layers (two layers of 16 filters of size $3\times3$ and two layers of 16 filters of size $3\times1$) with max pooling, followed by three dense layers (256 units, 32 units and 1 unit). 
All layers use a leaky rectifier as activation function with the exception of the output layer which uses the sigmoid function.
%specify the input features used for the CNN (or cross-reference to the appropriate subsection in Section 3 that describes them).

\begin{table}[t]

\caption{CNN architecture.}
\centering
\begin{tabular}{c|l}
          \textbf{Layer }& \textbf{Size}\\
          \hline
          Input &  time frames$\times$freq bands   \\ 
          Conv 1 & 16 (3$\times$3) filters \\
          Conv 2& 16 (3$\times$3) filters \\
          Conv 3& 16 (3$\times$1) filters \\
          Conv 4& 16 (3$\times$1) filters\\
          Dense 1& 256 units  \\
          Dense 2& 32 units \\
          Dense 3& 1 unit\\
		\end{tabular}
        \label{table:CNN_spectra_architecture}
\vspace{-3mm}        
\end{table}

\section{Evaluation}
\label{sec:results}
%to be written: what is the objective of the analysis, which data of the beehive will be used (insert a table), what we have obtained 
Several experiments are carried out using real-world audio data, with an aim to evaluate the performance of the proposed systems.

\subsection{Data}
Audio data from the NU-Hive \cite{cecchi2018a} project acquired in honey bee hives is used for training and evaluating the proposed system. 
In particular, the data from two hives has been used and for each hive a period of one day where the queen bee was present and one day without queen bee has been considered, for a total amount of 576 files of 10 min duration each ($\sim$ 96 hours). As reported in \cite{cecchi2018a}, the data was acquired continuously with $f_s=32$~kHz and the microphones are MEMS type positioned inside the hive, avoiding propolization. %Figure \ref{fig:hive} shows the bee colony and the installation of the acquisition system behind each hive.

%\begin{figure}[!b]
%    \centering
%    \includegraphics[width=6cm]{./img/hive.eps}
%    \caption{Bee colony used for the acquisition of the real data.}
%    \label{fig:hive}
%\end{figure}

\subsection{Experimental setup} \label{sec:expSetup}
Given the interest in constructing a system for a real-world scenario, we evaluate how well the classifiers are able to generalize to unseen hives. Thus, besides randomly splitting the dataset between train and test sets, we also implement a ``hive-independent'' splitting scheme. This means having training samples belonging only to certain hives, and testing using samples from other, unseen hives.

For the random scheme, a test size of 5\% of the total amount of data is used and, when applying the SVM classifier, all remaining data (95\%) is used in a single training set. For the CNN implementation, the remaining data is further split in half between the training and validation sets.
For the hive-independent scheme, the data is split in two, according to which hive they belong to; one is kept for training and the other for testing. In the case of the CNNs, a validation set, used for early stopping purposes, is obtained by randomly selecting 10\% of the training set data.

\subsection{Evaluation metrics}
The results of each experiment are evaluated using the area under the curve score (AUC) \cite{mesaros2018datasets}. Each experiment is run twice in different splits, following the same setup and parameters. We report the results on each run and the average AUC over the two.

\subsection{SVM Experiments}\label{sec:SVMexperiments}
Several experiments using SVMs with different sets of features are set up and run with the two split configurations reported in Section \ref{sec:expSetup}:

\begin{description}
\item[SVM$\_$MFCCs20:] each input sample is a vector of 20 MFCCs resulting from averaging the 10min MFCC coefficients over time.
\item[SVM$\_$HHTdwns20:]% for each 1s seconds frame for 10min recording, 20 features are extracted. 
the HHT spectrogram obtained for a 10min audio recording (see Section~\ref{sec:emd}) is aggregated over time and its maximum frequency limited to 6000~Hz. This process results in a vector with 6000 frequency bands representing one 10min recording. The frequency bands are further downsampled into 20 HHT bands, in order to reduce feature dimension.
\item[SVM$\_$MFCCs20$\_$HHTdwns20:] a combination of both representations described above is used. In total, a sample corresponding to a 10 min audio recording is represented by a vector of size 40 which is the concatenation of both feature vectors.
\item[SVM$\_$MEL120dwns20:] each Mel spectrogram of a 10 min audio recording is averaged over time resulting in a 120-dimensional vector which is further downsampled into 20 bands.
%\item[SVM$\_$stMFCCs20$\_$HHT:]
%\item[SVM$\_$MFCCs20$\_$HHT:]
\item[SVM$\_$LOG$\_$MEL120dwns20:] the log-Mel spectrograms of the 10 min audio recordings are averaged over time, and the frequency bands downsampled into 20 bands.
%\item[SVM$\_$stMEL120:]
\end{description}

\subsection{CNN Experiments}\label{sec:CNNexperiments}
Similar to the SVM experiments, the designed CNN model is trained with different features with both random and ``hive-independent" splits of the data.
As described in Section~\ref{sec:ml}, the proposed CNN approach uses spectrograms of 1 min audio data as input. These are computed by processing the audio with a sampling rate of 22.05~kHz, and applying a window size of 2048 samples and hop length of 512 samples. After, the resulting spectrograms are further transformed, as described in Section~\ref{sec:ml}, to highlight long-term contextual aspects. In specific, the spectrogram is segmented along the time dimension into 30 slices, each containing approximately 86 time frames ($\sim$2sec). The slices are further averaged over time and stacked together creating a matrix with 30 columns and the same original number of frequency bands.
The experiments are: 
\begin{description}
\item[CNN$\_$MFCCs20:]Uses as input data 20 MFCCs.
\item[CNN$\_$MEL120:]Input samples are obtained from Mel spectra with 120 frequency bands.
\item[CNN$\_$LOG$\_$MEL120:] This configuration uses the log-Mel spectra computed with 120 frequency bands.
\end{description}

The network described in Table~\ref{table:CNN_spectra_architecture} is trained over 100 epochs on batches of 145 samples with the RMSprop \cite{Goodfellow-et-al-2016} optimizer. Early stopping with a patience value of five and dropout of 50\% in the three last layers is employed during training.

\begin{figure}[!t]
  \centering
  \label{fig:SVMres}{\includegraphics[width=8.4cm]{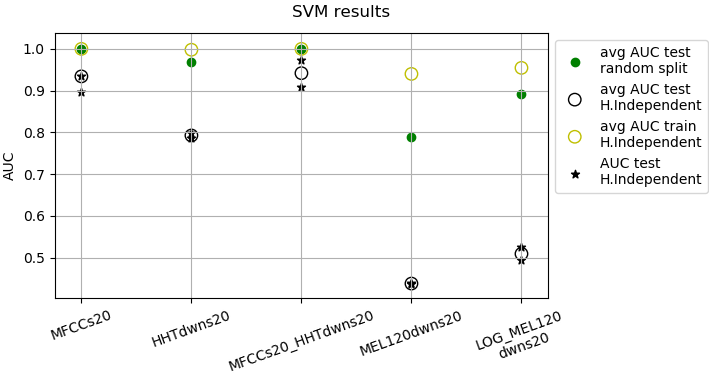}}
 \caption{[SVM results]{ The \boldmath$\star$ represents the AUC score on the test set for each fold of the hive-independent setup and SVM experiment. The \textcolor{yellow}{\(  \circ  \)} and \textbf{\(  \circ  \)} represent the average AUC score over the two folds in both train and test sets, respectively. The \textcolor{green}{\textbullet} reports the average AUC score of the test sets across the two folds of the random split setup. }}
\end{figure}

\begin{figure}
  \centering
  \label{fig:CNNres}{\includegraphics[width=8.4cm]{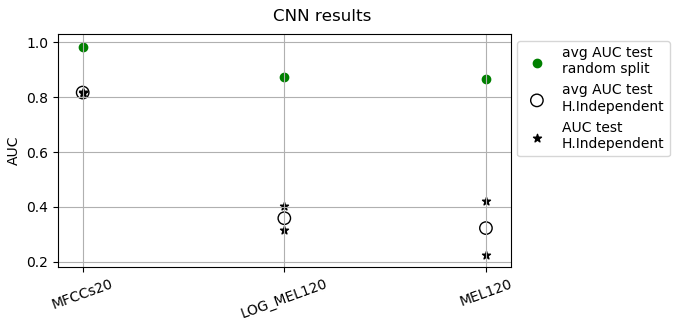}}
 \caption{[CNN results]{ The \boldmath$\star$ represents the AUC score on the test set for each fold of the hive-independent setup and CNN experiment. The \textbf{\( \circ \)} represents the average AUC score on test sets over the two folds and the \textcolor{green}{\textbullet} reports the average AUC score of the test sets across the two folds of the random split setup. }}
 \vspace{-4mm}
\end{figure}

\subsection{Results}
% since CNN and SVM results won't be exactly comparable due to the different data used, i've split the results in two tables
% not reportng the AUC in each fold for the Random split case, it was a question of focusing on hive -independent and to have clearer tbles.
%This section presents the main results from the experiments defined in  Sections \ref{sec:SVMexperiments} and \ref{sec:CNNexperiments}. %are presented in Figures \ref{fig:SVMres} and \ref{fig:CNNres} respectively.
\subsubsection{SVM results} %COMMENTED OUT FOR SPACE
The results of the SVM experiments are reported in Fig.~3 with the average AUC scores of the test sets over two folds in both the random split and the ``hive-independent" split setups. Additionally, the individual AUC scores of the test set in the ``hive independent" split setup are also included.
Observing the results for the averaged test AUC score in the random split setup ( \textcolor{green}{\textbullet} ), they are consistent with the reported results in \cite{Inolasco_dissertation} that indicate a perfect classification when using SVMs with RBF kernels and MFCCs on a random split setup. Here we further conclude that also the HHTs are good representations that yield similar classification results in this setup.

Given this, and as indicated in \cite{Inolasco_dissertation} and \cite{inolasco_DCASE2018}, the challenge when working with beehive sounds through a machine learning approach lies on learning classifiers that are able to generalize to different hives. This aspect is confirmed here with the reported lower values of the averaged test AUC scores in the ``hive-independent" setup  (\textbf{\(  \circ  \)}) for every experiment conducted. Despite the lower scores for the ``hive-independent" setup when compared with random splits, results indicate that the SVMs are successful in generalizing to unseen hives for most feature combinations.

An interesting result is presented in Fig.~3 \textbf{[MFCCs20$\_$ HHTdwns20]} where the averaged test AUC score ($\sim0.94$) in this setup  is better than the one reported in \textbf{[MFCCs20]}  ($\sim0.91$). These results indicate that the combination of these two features results in better predictions in a ``hive-independent" setup than the predictions carried out with classifiers learning from each feature individually.
It is also of notice in \textbf{[MEL120dwns20]} and \textbf{[LOG$\_$MEL120dwns20]}
the inadequacy of using these representations together with SVMs for this classification task on the ``hive-independent" setup.

\subsubsection{CNN results} %COMMENT OUT FOr SPACE
The resulting AUC scores for the experiments with CNNs are shown in Fig.~4. For the three CNN experiments carried out in a random split setup, the averaged AUC scores (\textcolor{green}{\textbullet}) show the ability of the procedure in this classification task for both  MFCCs and Mel spectra. %in particular when using MFCCs as input features. 
The challenge in predicting the beehive state in hives different than the ones where the model was trained is evident when observing the resulting averaged test AUC score in a ``hive-independent" setup, (\textbf{\(  \circ  \)}); overall the results in this setup decrease when comparing with the random split. Once again, as Fig.~4 \textbf{[MFCCs20]} shows, the use of MFCCs as feature is especially useful in this problem, while Mel spectra when used as features do not appear to generalize well to unseen hives given the resulting AUC lower than 0.5.

\vspace{-.5em}
\section{Conclusions}
\label{sec:conc}

% Focus discussion on Hive_independent setup , cite master thesis or dcase on how difficult this was..and how this is useful in a real scenario where we would deploy a classifier previously trained.

% 
This work explored the potential of traditional and neural network-based machine learning methods exploiting MFCCs, Mel spectrograms, and the HHT as feature extractors to determine the presence of the queen bee inside the hive. Several experiments on a real-world scenario have demonstrated the potential of the work exploiting the use of HHT as spectral representations and the design of representations suitable for modelling long-term temporal context. 
% EB: perhaps discuss a bit the results overall? The SVM seems to generalize well to unseen hives. The CNNs do not appear to generalize that well to unseen hives, although report good results in a hive-dependent scenario.
Results using SVMs show their ability to generalize to unseen hives and also the dominance of MFCCs as representations of the data when compared to other features. Better results were obtained when combining HHTs and MFCCs, which is an interesting point for further investigation.
The CNNs do not appear to generalize as well to unseen hives in the tested configurations, however they achieve good results in a hive-dependent scenario which indicates the feasibility of the application of deep learning methods to this unique problem, at least in a more controlled supervised scenario.

%ines - mention the need for evaluation in more hives!
Future work will further evaluate the methods in a hive independent scenario, for which the dataset must be augmented with new hives. A deeper investigation of CNNs in combination with the HHT and MFCC features is planned; we will also investigate the application of this procedure to the identification of other states of the honey bee hive, including swarming or pest presence.
The dataset\footnote{\url{https://zenodo.org/record/2563940\#.XGVwpDP7SUk}} and python code\footnote{\url{https://github.com/madzimia/Audio\_based\_identification\_beehive\_states.git}} developed for this work are publicly available.

\clearpage

\bibliographystyle{IEEEbib}
\bibliography{strings,refs}

\begin{thebibliography}{10}

\bibitem{Cane2007}
Alexandra-Maria Klein, Bernard~E Vaissi{\`e}re, James~H Cane, Ingolf
  Steffan-Dewenter, Saul~A Cunningham, Claire Kremen, and Teja Tscharntke,
\newblock ``Importance of pollinators in changing landscapes for world crops,''
\newblock {\em Proceedings of the Royal Society of London B: Biological
  Sciences}, vol. 274, no. 1608, pp. 303--313, 2007.

\bibitem{ntalampiras2012acoustic}
S.~Ntalampiras, I.~Potamitis, and N.~Fakotakis,
\newblock ``Acoustic detection of human activities in natural environments,''
\newblock {\em Journal of the Audio Engineering Society}, vol. 60, no. 9, pp.
  686--695, 2012.

\bibitem{Frings:1957}
H.~Frings and F.~Little,
\newblock ``Reactions of honey bees in the hive to simple sounds,''
\newblock {\em Science}, pp. 122--125, 1957.

\bibitem{Michelsen:1986}
A.~Michelsen, W.~H. Kirchner, and M.~Lindauer,
\newblock ``Sound and vibrational signals in the dance language of the
  honeybee, apis mellifera,''
\newblock {\em Behavioral Ecology and Sociobiology}, vol. 18, no. 3, pp.
  207--212, Jan. 1986.

\bibitem{Kirchner1993}
W.~H. Kirchner,
\newblock ``Acoustical communication in honeybees,''
\newblock {\em Apidologie}, vol. 24, no. 3, pp. 297--307, 1993.

\bibitem{Hrncir2006}
M.~Hrncir, F.~G. Barth, and J.~Tautz,
\newblock ``Vibratory and airborne sound-signals in bee communication,''
\newblock in {\em In Insect Sounds and Communication: Physiology, Behaviour,
  Ecology, and Evolution}, S.~Drosopoulos and M.~Claridge, Eds., pp. 421--436.
  CRC Press, 2006.

\bibitem{Hunt2013}
J.~H. Hunt and F.~J. Richard,
\newblock ``Intracolony vibroacoustic communication in social insects,''
\newblock {\em Insectes Sociaux}, vol. 60, pp. 403--417, 2013.

\bibitem{Cejrowski2018}
T.~Cejrowski, J.~Szyma{\'{n}}ski, H.~Mora, and D.~Gil,
\newblock ``Detection of the bee queen presence using sound analysis,''
\newblock in {\em Intelligent Information and Database Systems}, N.~T. Nguyen,
  D.~H. Hoang, T.-P. Hong, H.~Pham, and B.~Trawi{\'{n}}ski, Eds., pp. 297--306.
  Springer International Publishing, 2018.

\bibitem{robles2018}
A.~Robles, T.~Saucedo-Anaya, E.~González-Ramírez, and C.~Galván~Tejada,
\newblock ``Frequency analysis of honey bee buzz for automatic recognition of
  health status: A preliminary study,''
\newblock {\em Research in Computing Science}, vol. 142, pp. 89--98, June 2017.

\bibitem{Dietlein:1985}
D.G. Dietlein,
\newblock ``A method for remote monitoring of activity of honeybee colonies by
  sound analysis,''
\newblock {\em Journal of Apicultural Research}, vol. 24, no. 2, pp. 176--183,
  1985.

\bibitem{Ferrari:2006}
S.~Ferrari, M.~Silva, M.~Guarino, and D.~Berckmans,
\newblock ``Monitoring of swarming sounds in bee hives for prevention of honey
  loss,''
\newblock in {\em International Workshop on Smart Sensors in Livestock
  Monitoring}, Sept. 2006.

\bibitem{Ferrari:2008}
S.~Ferrari, M.~Silva, M.~Guarino, and D.~Berckmans,
\newblock ``Monitoring of swarming sounds in beehives for early detection of
  the swarming period,''
\newblock {\em Computers and Electronics in Agriculture}, vol. 65, pp. 72--77,
  2008.

\bibitem{quandour2014}
A.~Qandour, I.~Ahmad, D.~Habibi, and M.~Leppard,
\newblock ``Remote beehive monitoring using acoustic signals,''
\newblock {\em Acoustics Australia / Australian Acoustical Society}, vol. 42,
  no. 3, pp. 204--209, Dec. 2014.

\bibitem{Bromenshenk:2007}
J.~J. Bromenshenk,
\newblock ``Honey bee acoustic recording and analysis system for monitoring
  hive health,'' 2007,
\newblock US Patent 7549907.

\bibitem{chapter11_book}
Dan Stowell,
\newblock ``Computational bioacoustic scene analysis,''
\newblock in {\em Computational Analysis of Sound Scenes and Events},
  T.~Virtanen, M.~D. Plumbley, and D.~P.~W. Ellis, Eds., pp. 303--333.
  Springer, 2018.

\bibitem{Amlathe2018}
P.~Amlathe,
\newblock ``Standard machine learning techniques in audio beehive monitoring:
  Classification of audio samples with logistic regression, {K}-nearest
  neighbor, random forest and support vector machine,''
\newblock M.S. thesis, Utah State University, 2018.

\bibitem{Inolasco_dissertation}
I.~Nolasco,
\newblock ``Audio-based beehive state recognition,''
\newblock M.S. thesis, Queen Mary University of London, 2018.

\bibitem{dcase2017web_acousticScene_results}
``{DCASE Challenge 2017, Task1 results},''
  \url{http://www.cs.tut.fi/sgn/arg/dcase2017/challenge/task-acoustic-scene-classification-results}.

\bibitem{inolasco_DCASE2018}
I.~Nolasco and E.~Benetos,
\newblock ``{To bee or not to bee: Investigating machine learning approaches
  for beehive sound recognition},''
\newblock in {\em 2018 Workshop on Detection and Classification of Acoustic
  Scenes and Events (DCASE)}, 2018, pp. 133--137.

\bibitem{Bisot2017}
V.~Bisot, R.~Serizel, and S.~Essid,
\newblock ``Feature learning with matrix factorization applied to acoustic
  scene classification,''
\newblock {\em IEEE/ACM Transactions on Audio, Speech and Language Processing},
  vol. 25, no. 6, pp. 1216--1229, 2017.

\bibitem{huang2005}
N.~Huang,
\newblock ``Introduction to the {Hilbert-Huang} transform and its related
  mathematical problems,''
\newblock in {\em Hilbert-Huang Transform and Its Applications}, pp. 1--26.
  World Scientific, Sept. 2005.

\bibitem{cecchi2018a}
S.~Cecchi, A.~Terenzi, S.~Orcioni, P.~Riolo, S.~Ruschioni, and N.~Isidoro,
\newblock ``A preliminary study of sounds emitted by honey bees in a beehive,''
\newblock in {\em Audio Engineering Society Convention 144}, May 2018.

\bibitem{burton2018}
Jake Burton, David Frank, Mahdi Saleh, Nassir Navab, and Helen~L. Bear,
\newblock ``{The speaker-independent lipreading play-off; a survey of
  lipreading machines},''
\newblock in {\em IEEE International Image Processing Applications and Systems
  (IPAS)}, 2018.

\bibitem{chapter4_book}
R.~Serizel, V.~Bisot, S.~Essid, and G.~Richard,
\newblock ``Acoustic features for environmental sound analysis,''
\newblock in {\em Computational Analysis of Sound Scenes and Events},
  T.~Virtanen, M.~D. Plumbley, and D.~P.~W. Ellis, Eds., pp. 13--40. Springer,
  2018.

\bibitem{Huang903}
N.~E. Huang, Z.~Shen, S.~R. Long, M.~C. Wu, H.~H. Shih, Q.~Zheng, N.-C. Yen,
  C.~C. Tung, and H.~H. Liu,
\newblock ``The empirical mode decomposition and the {Hilbert} spectrum for
  nonlinear and non-stationary time series analysis,''
\newblock {\em Proceedings of the Royal Society of London A: Mathematical,
  Physical and Engineering Sciences}, vol. 454, no. 1971, pp. 903--995, 1998.

\bibitem{marple1999}
L.~Marple,
\newblock ``Computing the discrete-time "analytic" signal via fft,''
\newblock {\em IEEE Transactions on Signal Processing}, vol. 47, no. 9, pp.
  2600--2603, Sept 1999.

\bibitem{Abdelouahad20018}
A.~Abdelouahad, A.~Belkhou, A.~Jbari, and L.~Bellarbi,
\newblock ``Time and frequency parameters of semg signal — force
  relationship,''
\newblock in {\em 2018 4th International Conference on Optimization and
  Applications (ICOA)}, April 2018, pp. 1--5.

\bibitem{mesaros2018datasets}
A.~Mesaros, T.~Heittola, and D.~Ellis,
\newblock ``Datasets and evaluation,''
\newblock in {\em Computational Analysis of Sound Scenes and Events},
  T.~Virtanen, M.~D. Plumbley, and D.~Ellis, Eds., pp. 147--179. Springer,
  2018.

\bibitem{Goodfellow-et-al-2016}
Ian Goodfellow, Yoshua Bengio, and Aaron Courville,
\newblock {\em Deep Learning},
\newblock MIT Press, 2016,
\newblock \url{http://www.deeplearningbook.org}.

\end{thebibliography}

\end{document}